\documentclass[aps,prl,preprint,tightenlines,superscriptaddress,showpacs,byrevtex]{revtex4}

\usepackage{graphics}
\usepackage{dcolumn}  % Align table columns on decimal point
\usepackage{color}
\usepackage{epsf}

\graphicspath{{ps}}

\def\T3M{\tau \rightarrow 3\mu}

\def\E3P{\eta\rightarrow \pi^+\pi^-\pi^0}

\def \misP   {\not\!\! P}

\def \Btaunu {B^{-}\rightarrow\tau^{-}\overline{\nu}}
\newcommand{\lw}[1]{\smash{\lower1.7ex\hbox{#1}}}

\begin{document}

\preprint{\vbox{ \hbox{   }
                 \hbox{BELLE-CONF-0428}
                 \hbox{ICHEP04 11-0675}
%                 \hbox{hep-ex -----}
}}

\title{ \quad\\[0.5cm] 
       Search for $\Btaunu$ at Belle}

%%%% insert the authorlist here. BEFORE the abstract !!!
%%% Paper:    
%%% Journal:  summer 2004 conference papers (PRL format)
%%% Contacts: 
%%% Last revised on July 14, 2004 16:40:00 EDT
%%% Non-responding authors or those who said NO are commented out.
%%% ====================================================================
%%% Click the RELOAD button on your web browser to see the updated file.
%%% ====================================================================
%%% Use \input{author} to insert this material into your latex file.
%%%%% Force institutions to appear in alphabetical order when typeset.
\affiliation{Aomori University, Aomori}
\affiliation{Budker Institute of Nuclear Physics, Novosibirsk}
\affiliation{Chiba University, Chiba}
\affiliation{Chonnam National University, Kwangju}
\affiliation{Chuo University, Tokyo}
\affiliation{University of Cincinnati, Cincinnati, Ohio 45221}
\affiliation{University of Frankfurt, Frankfurt}
\affiliation{Gyeongsang National University, Chinju}
\affiliation{University of Hawaii, Honolulu, Hawaii 96822}
\affiliation{High Energy Accelerator Research Organization (KEK), Tsukuba}
\affiliation{Hiroshima Institute of Technology, Hiroshima}
\affiliation{Institute of High Energy Physics, Chinese Academy of Sciences, Beijing}
\affiliation{Institute of High Energy Physics, Vienna}
\affiliation{Institute for Theoretical and Experimental Physics, Moscow}
\affiliation{J. Stefan Institute, Ljubljana}
\affiliation{Kanagawa University, Yokohama}
\affiliation{Korea University, Seoul}
\affiliation{Kyoto University, Kyoto}
\affiliation{Kyungpook National University, Taegu}
\affiliation{Swiss Federal Institute of Technology of Lausanne, EPFL, Lausanne}
\affiliation{University of Ljubljana, Ljubljana}
\affiliation{University of Maribor, Maribor}
\affiliation{University of Melbourne, Victoria}
\affiliation{Nagoya University, Nagoya}
\affiliation{Nara Women's University, Nara}
\affiliation{National Central University, Chung-li}
\affiliation{National Kaohsiung Normal University, Kaohsiung}
\affiliation{National United University, Miao Li}
\affiliation{Department of Physics, National Taiwan University, Taipei}
\affiliation{H. Niewodniczanski Institute of Nuclear Physics, Krakow}
\affiliation{Nihon Dental College, Niigata}
\affiliation{Niigata University, Niigata}
\affiliation{Osaka City University, Osaka}
\affiliation{Osaka University, Osaka}
\affiliation{Panjab University, Chandigarh}
\affiliation{Peking University, Beijing}
\affiliation{Princeton University, Princeton, New Jersey 08545}
\affiliation{RIKEN BNL Research Center, Upton, New York 11973}
\affiliation{Saga University, Saga}
\affiliation{University of Science and Technology of China, Hefei}
\affiliation{Seoul National University, Seoul}
\affiliation{Sungkyunkwan University, Suwon}
\affiliation{University of Sydney, Sydney NSW}
\affiliation{Tata Institute of Fundamental Research, Bombay}
\affiliation{Toho University, Funabashi}
\affiliation{Tohoku Gakuin University, Tagajo}
\affiliation{Tohoku University, Sendai}
\affiliation{Department of Physics, University of Tokyo, Tokyo}
\affiliation{Tokyo Institute of Technology, Tokyo}
\affiliation{Tokyo Metropolitan University, Tokyo}
\affiliation{Tokyo University of Agriculture and Technology, Tokyo}
\affiliation{Toyama National College of Maritime Technology, Toyama}
\affiliation{University of Tsukuba, Tsukuba}
\affiliation{Utkal University, Bhubaneswer}
\affiliation{Virginia Polytechnic Institute and State University, Blacksburg, Virginia 24061}
\affiliation{Yonsei University, Seoul}
  \author{K.~Abe}\affiliation{High Energy Accelerator Research Organization (KEK), Tsukuba} % KEK
  \author{K.~Abe}\affiliation{Tohoku Gakuin University, Tagajo} % TohokuGakuin
  \author{N.~Abe}\affiliation{Tokyo Institute of Technology, Tokyo} % TIT
  \author{I.~Adachi}\affiliation{High Energy Accelerator Research Organization (KEK), Tsukuba} % KEK
  \author{H.~Aihara}\affiliation{Department of Physics, University of Tokyo, Tokyo} % Tokyo
  \author{M.~Akatsu}\affiliation{Nagoya University, Nagoya} % Nagoya
  \author{Y.~Asano}\affiliation{University of Tsukuba, Tsukuba} % Tsukuba
  \author{T.~Aso}\affiliation{Toyama National College of Maritime Technology, Toyama} % Toyama
  \author{V.~Aulchenko}\affiliation{Budker Institute of Nuclear Physics, Novosibirsk} % BINP
  \author{T.~Aushev}\affiliation{Institute for Theoretical and Experimental Physics, Moscow} % ITEP
% ...
% Lausanne
\collaboration{The Belle Collaboration}

%\collaboration{Belle Collaboration}
\noaffiliation

\begin{abstract}
% abstract goes here

We present a search for the decay $\Btaunu$ in a $140~\textrm{fb}^{-1}$
data sample collected at the $\Upsilon(4S)$ resonance with the Belle detector 
at the KEKB asymmetric $B$ factory. 
Combinatorial and continuum backgrounds are suppressed by selecting a sample
of events with one fully reconstructed $B$. The decay products of the other 
side $B$ in the event are analyzed to search for a $\Btaunu$ decay.
We find no significant evidence for a signal and set a 90\% confidence level
upper limit of $
Br(B^{-}\rightarrow\tau^{-}\overline{\nu}) < 2.9\times 10^{-4}$.
All results are preliminary.

\end{abstract}
\pacs{}  

\maketitle

\tighten

{\renewcommand{\thefootnote}{\fnsymbol{footnote}}}
\setcounter{footnote}{0}

%===========================================================================
%\section{Introduction}
%\label{sec:intro}

The purely leptonic decay $B^{-}\rightarrow\ell^{-}\overline{\nu}$ 
is of particular interest
since it provides direct measurement of the product of 
the Cabibbo-Kobayashi-Maskawa(CKM) matrix
element $V_{ub}$ and the $B$ meson form factor $f_{B}$.
In the Standard Model(SM), the branching fraction of the decay 
$B^{-}\rightarrow\ell^{-}\overline{\nu}$ is given as
\begin{equation}
Br(B^{-}\rightarrow\ell^{-}\overline{\nu}) = \frac{G_{F}^{2}m_{B}m_{\ell}^{2}}{8\pi}\left(1-\frac{m_{\ell}^{2}}{m_{B}^{2}}\right)^{2}f_{B}^{2}|V_{ub}|^{2}\tau_{B}
\end{equation}
where $G_{F}$ is the Fermi coupling constant, $m_{\ell}$ and $m_{B}$ are
the charged lepton and $B$ meson masses, $\tau_{B}$ is the $B^{-}$ lifetime.
The dependence of the lepton mass arises from helicity conservation and
helicity suppresses the muon and electron channels.

In the extension of the Standard Model, one expects significant modification to
the $\Btaunu$ decay branching fraction.
In the two-Higgs doublet model, the decay can occur via a charged 
Higgs particle. The $\Btaunu$ branching fraction is given as
\begin{equation}
Br(\Btaunu)
 =  \frac{G_{F}^{2}m_{B}m_{\tau}^{2}}{8\pi}\left(1-\frac{m_{\tau}^{2}}{m_{B}^{2}}\right)^{2}f_{B}^{2}|V_{ub}|^{2}\tau_{B}\times r_{H},
\end{equation}
where $r_{H}$ is defined as
\begin{equation}
r_{H} = \left(1-\frac{\tan^{2}\beta}{1+\tilde{\epsilon_{0}}\tan\beta}\frac{m_{B}^{2}}{m_{H}^{2}}\right)^{2}
\label{eq:r_H}
\end{equation}
and $\tan\beta$ is the ratio of vacuum expectation values of two Higgs 
doublets \cite{Hou:1992sy}. Once we get an upper limit on $Br(\Btaunu)$,
we can give a constraint on $\tan\beta$ and $m_{H}$.
Similarly, in $R$-parity violating extensions of the MSSM, $\Btaunu$
may be mediated by scalar supersymmetric particles\cite{Baek:1999ch}.
Hence, upper limits on the $\Btaunu$ branching fraction constrain $R$-parity
violating couplings.

%No observation of a $\Btaunu$ signal has been reported yet in the literature.
No evidence for an enhancement relative to the Standard Model prediction
was observed in previous experimental studies by 
CLEO\cite{Artuso:1995ar, Browder:2000qr}, ALEPH\cite{Buskulic:1994gj},
L3\cite{Acciarri:1996bv}, DELPHI\cite{Abreu:1999xe} and BABAR\cite{Aubert:2004kz}.
The most stringent upper limit has been achieved by the BABAR Collaboration : 
$Br(\Btaunu) < 4.2\times 10^{-4}$ at $90\%$ C.L. from a sample of fully reconstructed 
$B$ and semi-leptonic decays.

%===========================================================================
%\section{Data Sample}
%\label{sec:data_sample}

We use a $140~\textrm{fb}^{-1}$ data sample containing 
$152.0\times 10^{6}$ $B$ meson pairs collected with the Belle detector
at the KEKB asymmetric energy $e^{+}e^{-}$ ($3.5$ on $8$ GeV) collider
\cite{Kurokawa:2003} operating at the $\Upsilon(4S)$ resonance 
($\sqrt{s} = 10.58$ GeV). 
%The sample contains $140.0\times 10^{6}$ produced $B\overline{B}$ pairs. 
The Belle detector is a large-solid-angle
magnetic spectrometer consisting of a three-layer silicon vertex detector,
a $50$-layer central drift chamber (CDC), a system of aerogel threshold
$\check{\textrm{C}}$erenkov counters (ACC), time-of-flight scintillation 
counters (TOF), and an electromagnetic calorimeter comprised of
CsI(Tl) crystals (ECL)  
located inside a superconducting solenoid coil that provides a $1.5$ T 
magnetic field. An iron flux-return located outside of the coil is 
instrumented to identify $K_{L}^{0}$ and muons. 
The detector is described in detail elsewhere \cite{belle_detector:2003}.

Fully reconstructed $B$ mesons, $B_{rec}$, are observed in the following decay modes:  
$B^{+}\rightarrow\overline{D}^{(*)0}\pi^{+}$, $\overline{D}^{(*)0}\rho^{+}$, 
$\overline{D}^{(*)0}a_{1}^{+}$ and $\overline{D}^{(*)0}D_{S}^{(*)+}$.
$\overline{D}^{0}$ candidates are reconstructed as 
$\overline{D}^{0}\rightarrow K^{+}\pi^{-}$, $K^{+}\pi^{-}\pi^{0}$,
$K^{+}\pi^{-}\pi^{+}\pi^{-}$, $K_{s}^{0}\pi^{0}$, $K_{s}^{0}\pi^{-}\pi^{+}$,
$K_{s}^{0}\pi^{-}\pi^{+}\pi^{0}$ and $K^{-}K^{+}$.
$\overline{D}^{*0}$ mesons are reconstructed by combining the 
$\overline{D}^{0}$ candidates with a pion or a photon.
% removed due to redundancy:  $\overline{D}^{*0}\rightarrow \overline{D}^{0}\pi^{0}$ and $\overline{D}^{0}\gamma$.
$D_{S}^{+}$ candidates are reconstructed in the decay modes
$D_{S}^{+}\rightarrow K_{S}^{0}K^{+}$ and $K^{+}K^{-}\pi^{+}$, and
$D_{S}^{*+}$ mesons are reconstructed by combining the $D_{S}^{+}$
candidates with a photon.
% redundant with text:  $D_{S}^{*+}\rightarrow D_{S}^{+}\gamma$.
All the tracks and photon candidates in the event not used to reconstruct
the $B_{rec}$ are studied to search for 
$B^{-}\rightarrow\tau^{-}\overline{\nu}$.
The advantage of having a sample of fully reconstructed $B$ meson is to 
provide a strong suppression of the combinatorial and continuum background
events. The disadvantage is the low efficiency of full $B$ meson 
reconstruction (about $0.3\%$).
Charged $B$ pair events are generated from $\Upsilon(4s)$ resonance
$(\sqrt{s}\sim10.58~\textrm{GeV})$, where the $B^{+}$ or $B^{-}$ is generated with specific momentum and energy. 
Selection of the fully reconstructed $B$ candidates
is made according to the values of two variables :
the beam constraint mass 
$M_{bc}\equiv\sqrt{E_{beam}^{2} - p_{B}^{2}}$
and the energy difference $\Delta E\equiv E_{B} - E_{beam}$. 
Here, $E_{B}$ and $p_{B}$ are the reconstructed energy and momentum
of the fully reconstructed $B$ candidate in the center-of-mass (CM) system,
and $E_{beam}$ is the beam energy in the CM frame.
%\begin{equation}
%\Delta E = E_{B}^{cm} - E_{beam}^{cm}
%\end{equation}
%where $E_{B}^{cm}$ is the energy of the $B$ meson and $E_{beam}^{cm}$ is the
%beam energy, both in the $\Upsilon(4s)$ rest frame; $M_{bc}$, the beam 
%constraint mass, is defined as:
%\begin{equation}
%M_{bc} = \sqrt{(E_{beam}^{cm})^{2} - (P_{B}^{cm})^{2}} 
%\end{equation}
%where $P_{B}^{cm}$ is the measured momentum of the reconstructed $B$ meson
%in the CM frame.

\begin{figure}
\centerline{
\epsfxsize=7cm \epsfbox{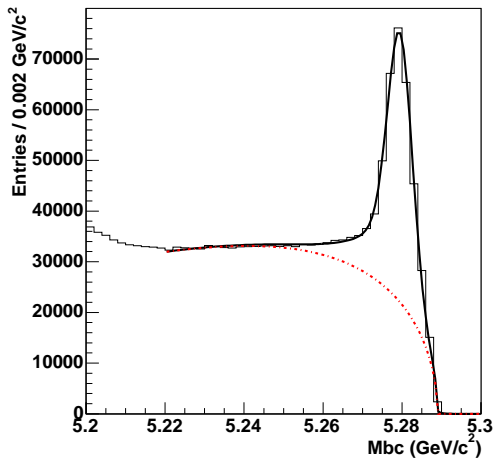} 
}
\caption{Distribution of the beam energy constrained mass in data for
	fully reconstructed $B$ mesons (histogram). The solid curve shows the 
	result of the fit and the background components as the dotted curve.}
    \label{mbc_fit_ver3_dE008}
\end{figure}

The $M_{bc}$ distribution of reconstructed $B$ candidates is fit with
the sum of an Argus function \cite{Albrecht:1986nr} 
and a Crystal Ball function \cite{Bloom:1983pc}.
The Argus function models the continuum and combinatorial background whereas
the Crystal Ball function models the signal component, which peaks at the
$B$ mass. The purity is defined as $S/(S+B)$, where $S~(B)$ is the number of 
signal (background) events for $M_{bc} > 5.27~\textrm{GeV}/c^{2}$, 
as determined from a fit. 
Figure \ref{mbc_fit_ver3_dE008} shows the $M_{bc}$ distribution 
for all $B_{rec}$ candidates in our data set.
The yield $N_{B^{+}B^{-}}$ of the sample containing one $B_{rec}$ is
determined as the area of the fitted Crystal Ball function. We obtain 
$N_{B^{+}B^{-}} = (2.40\pm 0.15)\times 10^{5}$ and $0.57$ of the purity, 
where the uncertainty on $N_{B^{+}B^{-}}$ is dominated by systematic errors.
We define the $B_{rec}$ signal region to be 
$-0.08 < \Delta E < 0.06~\textrm{GeV}$ and 
$M_{bc} > 5.27~\textrm{GeV}/c^{2}$.
In order to avoid experimenter bias, the signal region in data is blinded
until the selection criteria are finalized.

% --------------------------------------------------

%===========================================================================
%\section{Event Selection}
%\label{sec:event_selection}

Once a reconstructed $B$ candidate has been identified, $\Btaunu$ signal
candidates are selected by considering all tracks and clusters in the event
which are not used in the fully reconstructed $B$.
Candidate events are required to have one or three signal-side charged
track(s) with the total charge which is opposite that of the reconstructed $B$.
In the events where a $B_{rec}$ is reconstructed, we search for decays
into a $\tau$ plus a neutrino. The $\tau$ lepton is identified in the 
following decay channels: $\tau^{-}\rightarrow\mu^{-}\nu\bar{\nu}$,
$\tau^{-}\rightarrow e^{-}\nu\bar{\nu}$, 
$\tau^{-}\rightarrow\pi^{-}\nu$,
$\tau^{-}\rightarrow\pi^{-}\pi^{0}\nu$, and 
$\tau^{-}\rightarrow\pi^{-}\pi^{+}\pi^{-}\nu$.

We require the charged particles to be identified as leptons or pions.
For the lepton and single-pion modes we reject events with $\pi^{0}$ mesons
in the recoil against $B_{rec}$.
The event is required to have zero charge and $E_{ECL}$ less than a certain
value ($E_{ECL}<1.0~\textrm{GeV}$ for $\tau^{-}\rightarrow\mu^{-}\nu\bar{\nu}$
and $e^{-}\nu\bar{\nu}$, $E_{ECL}<1.2~\textrm{GeV}$ for
$\tau^{-}\rightarrow\pi^{-}\nu$ and $\pi^{-}\pi^{+}\pi^{-}\nu$ and
$E_{ECL}<2.2~\textrm{GeV}$ 
for $\tau^{-}\rightarrow\pi^{-}\pi^{0}\nu$)
where $E_{ECL}$ is defined as $E_{ECL} = E_{tot} - E_{B_{rec}} - E_{track}$,
the energy deposition in the ECL calorimeter.
An additional requirement $E_{\pi^{0}}/E_{ECL} > 0.2$ is applied
for $\tau^{-}\rightarrow\pi^{-}\pi^{0}\nu$.
Further requirements are made on the total momentum of the track(s) in the CM
frame ($p_{\pi^{-}} > 1.0~\textrm{GeV}/c$ for 
$\tau^{-}\rightarrow\pi^{-}\nu$,
$p_{\pi^{-}\pi^{+}\pi^{-}} > 1.2~\textrm{GeV}/c$ for
$\tau^{-}\rightarrow\pi^{-}\pi^{+}\pi^{-}\nu$),
the total missing momentum of the event ($\misP > 1.0~\textrm{GeV}/c$
for $\tau^{-}\rightarrow\pi^{-}\nu$ and $\pi^{-}\pi^{0}\nu$,
$\misP > 1.2~\textrm{GeV}/c$ for 
$\tau^{-}\rightarrow\pi^{-}\pi^{+}\pi^{-}\nu$) 
and the invariant mass of two or three pions
$0.55 < m_{\pi\pi} < 0.95~\textrm{GeV}/c^{2}$ and 
$1.0 < m_{\pi\pi\pi} < 1.4~\textrm{GeV}/c^{2}$.
The selection efficiencies for the $\tau$ decay channels we consider in this
analysis are determined from Monte Carlo simulations 
of $B^{-}\rightarrow\tau^{-}\bar{\nu}$  events.
We compute the efficiency as the ratio of the number of events surviving 
each of our selections over the number of fully reconstructed $B$.

%===========================================================================
%\section{Background Estimation}
%\label{sec:backgroud}

The expected background is composed of events from continuum and combinatorial
background, along with fully reconstructed $B$ meson events.
Backgrounds consists primarily of $B^{+}B^{-}$ events in which the fully
reconstructed $B$ has been correctly reconstructed. 
%but in which the
%accompanying $B$ decays to a lepton or pion(s) which are not reconstructed
%by the tracking detectors or calorimeter.
This ``peaking'' background is directly determined from Monte Carlo 
simulations of $B^{+}B^{-}$ events.
The continuum and combinatorial background is determined from the number
of events in $M_{bc}$ sideband data, scaled by the ratio of the areas 
of the fitted Argus function in the signal and sideband regions.
% due to the limited MC statistics. 
We fit $M_{bc}$ distributions after preselection and assume the ratio of 
the fitted Argus is unchanged after all selection criteria have been applied.
This assumption is consistent with the observed distributions in the Monte
Carlo as well as the data in $\Delta E$ sideband.

%==========================================================================
%\section{Systematic Uncertainties}
%\label{sec:systematic}

The main sources of uncertainty we consider in the determination of the
$Br(B^{-}\rightarrow\tau^{-}\overline{\nu})$ are 
uncertainty in the number of $B^{+}B^{-}$ events with one reconstructed $B$,
uncertainty in the determination of the signal efficiency and
uncertainty in the determination of the number of expected background events.
The number of $B^{+}B^{-}$ events is determined as the area 
of the Crystal Ball function fitted to the $M_{bc}$ distribution. 
Using a Gaussian function as an alternative fitting function,
we obtain a relative change in the number of events and
this difference is assigned as the systematic uncertainty on  
the number of $B^{+}B^{-}$ events.
A systematic error due to uncertainty in the amount of
neutral $B$ background in the number of $B^{+}B^{-}$ is also considered.
The main contribution to the systematic uncertainties in the determination 
of the efficiencies come from uncertainty on tracking efficiency, 
Monte Carlo statistics, $E_{ECL}$ energy and particle identification.
The uncertainty in the expected background comes from Monte Carlo statistics,
$E_{ECL}$ energy uncertainty and use of the Argus fit function.

%==========================================================================
%\section{Results}
%\label{sec:results}

\begin{figure}
\centerline{
\epsfxsize=8cm \epsfbox{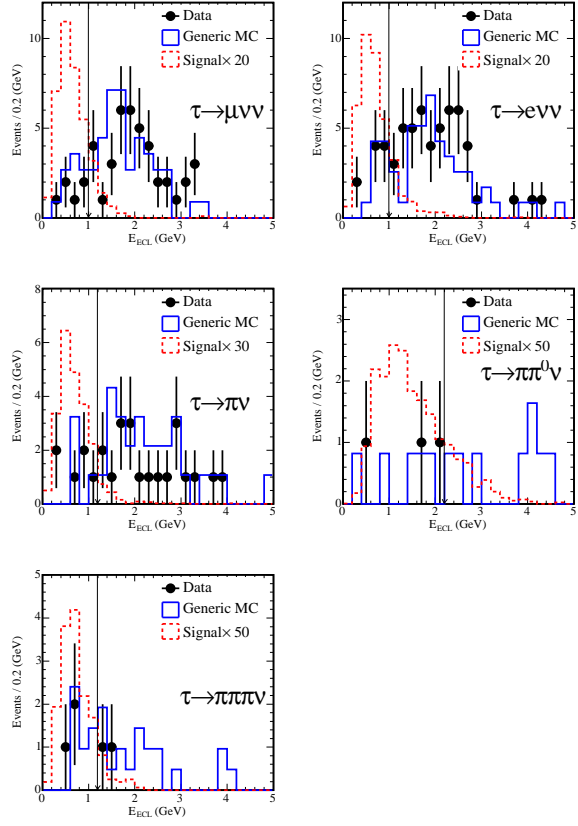} 
}
\caption{$E_{ECL}$ distributions in the data after
	all selection requirements except the one on $E_{ECL}$ 
	have been applied. The vertical arrow is the requirement
	on the $E_{ECL}$ in each selection.}
    \label{ecl_signal_mod}
\end{figure}

Figure \ref{ecl_signal_mod} shows the $E_{ECL}$ distributions in the data after
all selection requirements except the one on $E_{ECL}$ have been applied
compared with the expected background. Each distribution refers to 
a different selections and the plots show no evidence of signal in data.
%Figure \ref{mbc_de_box_open} shows the distributions of $\Delta E$ 
%{\it vs.} $M_{bc}$ in the data after all selection criteria have been applied.
We find a total of 28 candidates in the signal region where 
$33.5 \pm 5.0$ background events are expected. 
This uncertainty in the background expectation is due to both statistical and 
systematic errors.

% -------------------------------------------------

In order to extract the upper limit on the branching fraction for
$B^{-}\rightarrow\tau^{-}\overline{\nu}$, we combine the results of the 
different selections.
We use the likelihood ratio estimator, $Q$. Here, $Q$ is defined as
${\cal L}(s+b) / {\cal L}(b)$,
where ${\cal L}(s+b)$ and ${\cal L}(b)$ are the likelihood functions for 
signal plus background and background only hypotheses, respectively.
The likelihood functions ${\cal L}(s+b)$ and ${\cal L}(b)$ are defined as
\begin{equation}
{\cal L}(s+b) = \prod^{n_{ch}}_{i=1}
\frac{e^{-(s_{i}+b_{i})}(s_{i}+b_{i})^{n_{i}}}{n_{i}!}\textrm{, }
%\end{equation}
%\begin{equation}
~{\cal L}(b) = \prod^{n_{ch}}_{i=1}
\frac{e^{-b_{i}}b_{i}^{n_{i}}}{n_{i}!}
\end{equation}
where $n_{ch}$ is the number of selection channels, $s_{i}$ and $b_{i}$
are the expected number of signal and background events, respectively, and
$n_{i}$ is the number of observed events in each channel.
The number of signal events $s_{i}$ can be written as
$s_{i} = \varepsilon_{i}\cdot N_{B^{+}B^{-}}\cdot Br(B^{-}\rightarrow \tau^{-}\nu)$,
where $\varepsilon_{i}$ is the selection efficiency for the $i$-th channel
and $N_{B^{+}B{-}}$ is the number of $B^{+}B{-}$ events with one reconstructed
$B_{rec}$.
We set a $90\%$ C.L. upper limit using a simple Monte Carlo generating random
experiments for different values of the branching fraction
$Br(B^{-}\rightarrow \tau^{-}\nu)$. The confidence level for the signal
hypothesis can be written as
\begin{equation}
CL_{s} = \frac{CL_{s+b}}{CL_{b}} = \frac{N_{Q_{s+b\leq Q}}}{N_{Q_{b\leq Q}}} 
%CL_{s} = \frac{CL_{s+b}}{CL_{b}} = \frac{P_{s+b}(Q\leq Q_{obs})}{P_{b}(Q\leq Q_{obs})} 
\end{equation}
where $N_{Q_{s+b\leq Q}}$ and $N_{Q_{b\leq Q}}$ are the number of the generated
experiments which have a likelihood ratio less than or equal to the measured
one, in the signal plus background hypothesis and background only hypothesis,
respectively. The $90\%$ C.L. upper limit is obtained for $CL_{s} = 1 - 0.9$, 
where Gaussian systematic uncertainties are incorporated.
This ``Modified Frequentist approach'' or ``CLs method'' method is
described in detail in reference \cite{Junk:1999kv}.

\begin{figure}
\centerline{
\epsfxsize=8cm \epsfbox{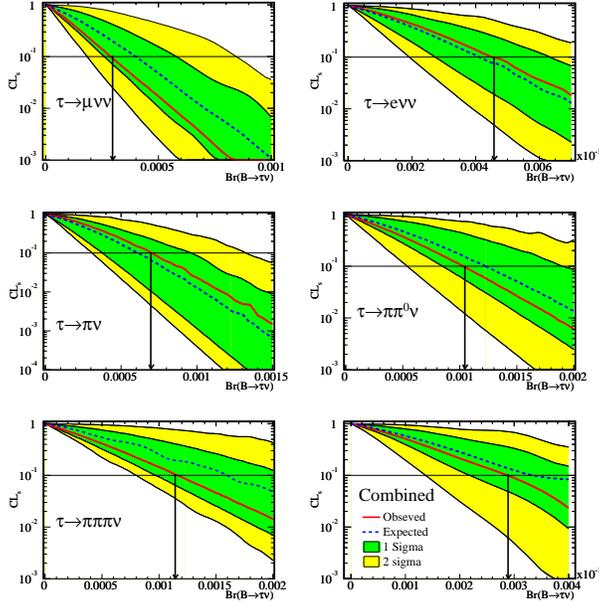} 
}
    \caption{The confidence level for the signal hypothesis $CL_{s}$
	 as a function of $Br(B^{-}\rightarrow \tau^{-}\nu)$ obtained 
	from the likelihood approach. The solid curves are the observed 
	results, the dashed curves the expected. The shaded areas represent
	the symmetric $1\sigma$ and $2\sigma$ bands. The intersections of the
	curves with the horizontal line at $CL_{s} = 0.1$ give the limits
	on $Br(B^{-}\rightarrow \tau^{-}\nu)$ at the 90\% confidence level.}
    \label{expected_limit_w_sys}
\end{figure}

Figure \ref{expected_limit_w_sys} shows 
the confidence level for the signal hypothesis $CL_{s}$
as a function of $Br(B^{-}\rightarrow \tau^{-}\nu)$ obtained 
from the likelihood approach. The solid curves are the observed 
results, the dashed curves the expected. The shaded areas represent
the symmetric $1\sigma$ and $2\sigma$ bands. Intersection of the
curves with the horizontal line at $CL_{s} = 0.1$ gives the limits
on $Br(B^{-}\rightarrow \tau^{-}\nu)$ at the 90\% confidence level, with
systematic uncertainties taken into account. 
The limits we observe lie within the $1\sigma$ expected regions.

% -------------------------------------------------

Table \ref{tab:observed_limit} shows the branching fraction times efficiency 
for signal, expected background, observed events and upper limit on 
the branching fraction for each $\tau$ decay mode. 
We obtain a combined limit on the branching fraction of
%  REDUNDANT with equation inline  at the 90\% confidence level of
\begin{equation}
Br(B^{-}\rightarrow\tau^{-}\overline{\nu}) < 2.9\times 10^{-4}
~\textrm{at the 90\% C.L.}
\end{equation}

% -------------------------------------------------
\begin{table}
 \begin{center}
  \renewcommand{\baselinestretch}{1.3}
   \begin{normalsize}
    \begin{tabular}{c|ccc|c} \hline
      \lw{$\tau$ Decay Mode} &\lw{Efficiency$\times BR (\%)$} 
	&Background &Observed     &Observed limit\\
    &   &Expected   &Events      &($90\%$C.L.) \\\hline \hline 

 $\tau^{-}\rightarrow\mu^{-}\nu\bar{\nu}$ 
          &$9.2 \pm 1.5$  &$9.8 \pm 2.9$  &$6$ &$3.0\times 10^{-4}$\\
 $\tau^{-}\rightarrow e^{-}\nu\bar{\nu}$ 
          &$8.8 \pm 1.5$ &$9.4 \pm 2.9$  &$10$ &$4.6\times 10^{-4}$\\
 $\tau^{-}\rightarrow\pi^{-}\nu\bar{\nu}$ 
          &$4.1 \pm 0.4$  &$5.4 \pm 2.1$  &$6$ &$7.2\times 10^{-4}$\\
 $\tau^{-}\rightarrow\pi^{-}\pi^{0}\nu$ 
          &$1.8 \pm 0.2$  &$4.1 \pm 1.6$  &$3$ &$10.5\times 10^{-4}$\\
 $\tau^{-}\rightarrow\pi^{+}\pi^{-}\pi^{+}\nu$ 
          &$1.6 \pm 0.2$  &$4.8 \pm 1.6$  &$3$ &$11.7\times 10^{-4}$\\	
\hline\hline
 \multicolumn{4}{c|}{Combined}   &$2.9\times 10^{-4}$\\ \hline
    \end{tabular}
    \caption{Branching fraction times efficiency for signal, expected 
	background, observed events and upper limit on the branching fraction
	for each $\tau$ decay mode.  The uncertainties on the efficiencies
	and the expected background are due to the combination of limited statistics 
      and systematic errors.}
   \label{tab:observed_limit}
  \end{normalsize}
 \end{center}
\end{table}

\begin{figure}
\centerline{
\epsfxsize=7cm \epsfbox{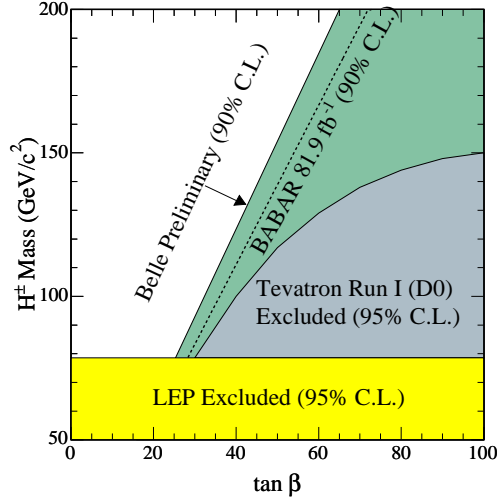} 
}
    \caption{The $90\%$ C.L. exclusion boundaries in the 
	$[M_{H^{+}}, \tan\beta]$ plane obtained from the observed upper 
	limit on $Br(B^{-}\rightarrow \tau^{-}\nu)$.}
    \label{Mh_tanb_exclude_expect}
\end{figure}

\begin{figure}
\centerline{
\epsfxsize=9cm \epsfbox{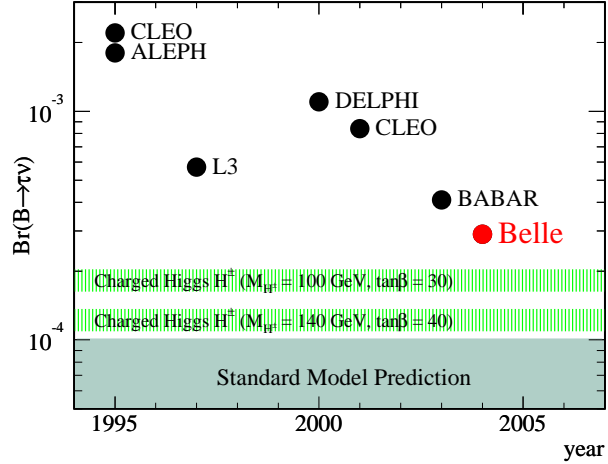} 
}
    \caption{Summary of searches for $\Btaunu$ and upper limits compared
	with the corresponding SM prediction and the branching fraction
	predicted with the charged Higgs boson in two parameter sets.}
    \label{Btaunu_history}
\end{figure}

In the two-Higgs doublet model, the branching fraction 
$Br(B^{-}\rightarrow \tau^{-}\nu)$ is enhanced by a factor of
$[1- (m_{B}/m_{H})^{2}\tan^{2}\beta]^{2}$ for $\tilde{\epsilon_{0}}=0$
in equation (\ref{eq:r_H}). With $m_{B}=5279~\textrm{MeV}/c^{2}$ and 
$Br(\Btaunu) = 7.5\times 10^{-4}$ from the Standard Model prediction,
we set the constraint 
\begin{equation}
\frac{\tan\beta}{M_{H^{\pm}}} < 0.33 ~(\textrm{GeV}/c^{2})^{-1}
\label{eq:limit_tan_H}
\end{equation}
from the obtained limits on the branching fraction of $\Btaunu$.
Figure \ref{Mh_tanb_exclude_expect} shows the $90\%$ C.L. exclusion 
boundaries in the $[M_{H^{+}}, \tan\beta]$ plane obtained from 
(\ref{eq:limit_tan_H}) compared with other experimental searches
at LEP\cite{Bock:2000gk} and at the Tevatron\cite{Abazov:2001md}.
The direct search for charged Higgs bosons at LEP gives the constraint
$M_{H^{\pm}}>78.6~\textrm{GeV}/c^{2}$.
Figure \ref{Btaunu_history} shows the summary of searches for $\Btaunu$ 
and upper limits obtained by other experiments compared with the corresponding 
SM prediction and the branching fraction predicted with the charged Higgs
boson in two parameter sets.
Our result is the most stringent on this process.

%==========================================================================
%\section{Summary}
%\label{sec:summary}

In conclusion, we have performed a search for 
the $B^{-}\rightarrow\tau^{-}\overline{\nu}$ decay in a fully reconstructed $B$
sample. The analysis uses the following $\tau$ decay channels:
$\tau^{-}\rightarrow\mu^{-}\nu\bar{\nu}$,
$\tau^{-}\rightarrow e^{-}\nu\bar{\nu}$, 
$\tau^{-}\rightarrow\pi^{-}\nu$,
$\tau^{-}\rightarrow\pi^{-}\pi^{0}\nu$, and 
$\tau^{-}\rightarrow\pi^{-}\pi^{+}\pi^{-}\nu$.
The results of the search in the different channels have been combined.
No signal is observed and an upper limit has been set :
\begin{equation}
Br(B^{-}\rightarrow\tau^{-}\overline{\nu}) < 2.9\times 10^{-4}~~(90\%~\textrm{C.L.}),
\end{equation}
which represents the most stringent upper limit on this process to date.

%==========================================================================
%\vspace{1.0cm}
%\hspace{-1.0cm}
%\begin{large}
%{\bf Acknowledgments\\}
%\end{large}

%***** Acknowledgments *****
% use these two starting with pub # 98
%----------- Long version, for most papers ----------- 
We thank the KEKB group for the excellent operation of the
accelerator, the KEK Cryogenics group for the efficient
operation of the solenoid, and the KEK computer group and
the National Institute of Informatics for valuable computing
and Super-SINET network support. We acknowledge support from
the Ministry of Education, Culture, Sports, Science, and
Technology of Japan and the Japan Society for the Promotion
of Science; the Australian Research Council and the
Australian Department of Education, Science and Training;
the National Science Foundation of China under contract
No.~10175071; the Department of Science and Technology of
India; the BK21 program of the Ministry of Education of
Korea and the CHEP SRC program of the Korea Science and
Engineering Foundation; the Polish State Committee for
Scientific Research under contract No.~2P03B 01324; the
Ministry of Science and Technology of the Russian
Federation; the Ministry of Education, Science and Sport of
the Republic of Slovenia;  the Swiss National Science Foundation; the National Science Council and
the Ministry of Education of Taiwan; and the U.S.\
Department of Energy.

%-------- Short version, if necessary, for PRL -----------
%We thank the KEKB group for the excellent operation of the
%accelerator, the KEK Cryogenics group for the efficient
%operation of the solenoid, and the KEK computer group and
%the NII for valuable computing and Super-SINET network
%support.  We acknowledge support from MEXT and JSPS (Japan);
%ARC and DEST (Australia); NSFC (contract No.~10175071,
%China); DST (India); the BK21 program of MOEHRD and the CHEP
%SRC program of KOSEF (Korea); KBN (contract No.~2P03B 01324,
%Poland); MIST (Russia); MESS (Slovenia); Swiss NSF; NSC and MOE
%(Taiwan); and DOE (USA).

% -------------------------------------------------

% -------------------------------------------------

%==========================================================================

\end{document}